\documentclass[11pt,letterpaper]{JHEP3}
\usepackage[parfill]{parskip}    % Activate to begin paragraphs with an empty line rather than an indent
\usepackage{graphicx}
\usepackage{amssymb}
\usepackage{bbm}
\newcommand{\tr}{\mathrm{tr}}
\newcommand{\nn}{\nonumber}

\title{The B-model on the A-model NS5-brane} 
\author{Viktor Bengtsson, Martin Cederwall\\
  Fundamental Physics\\
  Chalmers University of Technology\\
  SE-412 96 G\"oteborg, Sweden\\
\email{viktor.bengtsson@chalmers.se, martin.cederwall@chalmers.se}}
\abstract{We formulate the dynamics for the NS5-brane of the A-model,
  via the `maximal form' method, which couples all background fields
  to the world-volume. This procedure provides the extra one- and
  five-forms fields of the extended Hitchin model. 
The generalised B-model emerges as the
  world-volume theory of this brane. The starting point of the
  construction is an embedding of the brane in a superspace geometry
  with 16 fermionic directions. The correspondence with Hitchin's
  formulation in terms of pure spinors is explained.}

\begin{document}

%\maketitle

\section{Introduction}

It has been conjectured that the topological string theories in six
dimensions \cite{Witten1,Witten2,Neitzke:2004ni}, 
the A-model and the B-model, are dual to each
other by ``S-duality'', where the two models are formulated on the
same manifold. The duality demands the presence of various branes in the
models \cite{Nekrasov:2004,Neitzke:2004pf}. In a similar spirit, it has been
demonstrated that the topological B-model is related to
an extended Hitchin model \cite{Hitchin:2004ut}, {\it i.e.}, a model describing
deformations of the generalised complex structure \cite{Pestun:2005rp}. The
extended Hitchin models contain field strengths that are even forms for the
A-model and odd forms for the B-model, and in addition obey certain
non-linear relations that most elegantly are expressed as pure spinor
conditions. D-branes in topological string theory have been considered
{\it e.g.} in refs. \cite{KapustinOrlov,KapustinLi,Chiantese}.

It has been argued that one way of obtaining the duality is to
consider a space-filling brane in the A-model, an NS5-brane (argued to
exist in \cite{Nekrasov:2004}). The theory on this brane should provide the
dynamics of the B-model. In ref. \cite{Bao:2006} (see also
\cite{Marino:1999af}) the topological M5-brane was directly dimensionally
reduced to the A-model NS5-brane, and it was shown that the equation
of motion for the (non-linearly) self-dual 2-form on the brane reduced
to the Kodaira--Spencer equation, governing the deformations of the
complex structure. It has also been demonstrated that the A-model is
described by a space-filling brane in the B-model
\cite{Dijkgraaf:2002ac,Kapustin:2004jm}. 

Our previous results \cite{Bao:2005,Bao:2006} 
have been formulated in a framework where the
topological models (topological string theories or M-theory
\cite{Dijkgraaf:2004te,Bonelli:2005rw}) are 
embedded directly in space-time supersymmetric versions of string or
M-theory, using a 6- or 7-dimensional supergravity with 16 real
supercharges on manifolds with $SU(3)$ or $G_2$ holonomy\footnote{The 
correspondence of this formulation with the relevant Hitchin models
will be the subject of a forthcoming publication \cite{toappear}}. This
formulation has the obvious advantage that the correspondence between
supersymmetric branes in the topological 
theories and brane instantons in the full
string or M-theory becomes direct. 

The purpose of the present paper is to demonstrate how the
string/5-brane duality naturally yields the extended Hitchin
models and generalised complex geometry. Our earlier space-time
supersymmetric methods are generalised to a setting where the full
coupling of the NS5-brane to the background fields are included. This
is done using the methods of
refs. \cite{Cederwall:1998a,Townsend:1997,Cederwall:1997,Bengtsson:2004,Cederwall:1998b},
where a 
world-volume gauge potential is introduced for every background field
strength. In such a formalism, successfully applied to a variety of
branes in string theory and M-theory, the number of world-volume
degrees of freedom generically is too high, and has to be reduced by
some self-duality relation consistent with the background couplings.
Thanks to the correspondence between self-duality and pure spinors in
$d=6$, this approach will turn
out to provide the correct fields to describe the extended Hitchin
model. In addition to these fields, the NS5-brane theory will itself
contain branes, namely the boundaries of A-model D-branes ending on
the NS5-brane. These will provide the D-branes of the B-model.

The earlier description of the 5-brane in topological M-theory, and of
the A-model NS5-brane, involves a non-linearly self-dual 2-form on the
brane. This corresponds to the field content in the ``unextended''
Hitchin model \cite{Hitchin:2000jd}, and was shown to produce
Kodaira--Spencer theory for the deformation of the complex structure. 
In order to match the extended Hitchin B-model \cite{Hitchin:2004ut}, the
world-volume theory must include a scalar and a 4-form, and the entire
field strength, which is composed by a 1-form, a 3-form and a 5-form,
has to obey some relation that on one hand reduces to the above
self-duality when the 1-form and 5-form vanish, on the other hand is
equivalent to a pure spinor condition. Only in $d=6$ does the
number of components in a self-dual odd (or even) form match the
number of components of a pure spinor of $SO(d,d)$. The correspondence
between the two parametrisations of the fields will be demonstrated.

Some words on what is done in this paper, and what is not.
We introduce superspace field strengths and Bianchi identities in the
$d=6$, $N=(1,1)$ superspace relevant for the A-model. The Bianchi
identities essentially follow from the basic Fierz identities, and
imply (with some extra assumptions) the form of the modified Bianchi
identities for the world-volume fields. A generic Ansatz is made for
the action of the NS5-brane, and its exact form is determined. The
calculations are quite involved, but are simplified to some extent by
the observation that we can manifest an $so(6)\oplus so(6)$ subalgebra
of $so(6,6)$. We find the explicit form of the action and of the
self-duality relations, and demonstrate that the action is
$\kappa$-symmetric ($\kappa$-symmetry and self-duality, as usual, go
hand in hand---it is the presence of the former, seen as a chirality
of the physical fermions, that allows for the decoupling and
consistent chiral truncation in the bosonic sector expressed by the
latter). The calculations rely on an intricate and very non-trivial
interplay between the 
background supergeometry, non-linear self-duality of world-volume
fields and $\kappa$-symmetry. 
What remains to be done, in order for the brane to describe a
topological model, is to identify the BRST operator $Q$ as a singlet
supersymmetry when holonomy is reduced to $SU(3)$, and consider the
stability relation imposed on the brane by demanding that it is
$Q$-invariant. This is the procedure performed in ref. \cite{Bao:2006} for the
reduction of the
topological M5-brane. We are confident that it will pose no problems,
and provide the correct generalisation of the Kodaira--Spencer
equation obtained there. Considering the simple form of projection
matrix, we expect the calculation to be straightforward.
The issue will be addressed in a
forthcoming publication.

%\cite{Bao:2005,deBoer:2005}

%As the analysis of this paper passes through two lifts of tensors to
%higher symmetries, the various index conventions are here summarised
%for later reference. 
%\begin{center}
%\begin{tabular}{llccc}
%Index type & Index letter & Range & Repr. & Group \\ \hline
%\bf{Bosonic} & & & & \\
%\footnotesize{Curved target space} & $\mu, \nu, \rho, ...$  & $1,
%..., 6$ & - & so(6) \\ 
%\footnotesize{Flat target space} & $a,b,c, ...$ & $1, ..., 6$ & - & so(6) \\
%\footnotesize{Curved world volume} & $i,j,k, ...$ & $1, ..., 6$ & - & so(6) \\
%\bf{Fermionic} & & & & \\
%\footnotesize{Fermionic} & $\alpha, \beta, \gamma, ...$ & $1, ..., 8$
%& $\mathbf{8}$ & so(6) \\ 
%\footnotesize{Pre-lift fermionic} & $A,B,C, ...$ & $1,...,4$ &
%$\mathbf{4} \oplus \bar{\mathbf{4}}$ & so(6) \\ 
%\footnotesize{Lifted fermionic} & $\! \! \! \! \left\{ \!
%  \! \begin{array}{l} A,B,C, ... \\ A',B',C', ... \end{array} \right.$
%& $1, ..., 4$& $(\mathbf{4},\mathbf{4}) \oplus
%(\bar{\mathbf{4}},\bar{\mathbf{4}})$ & so(6)  $\! \oplus \!$ so(6) 
%\end{tabular}
%\end{center}

A note on index conventions: We use $a,b,\ldots$ for background
Lorentz indices, and $i,j,\ldots$ for curved world-volume
indices. Fermions transform in $(4\oplus\bar4,2)$ of $so(6)\oplus
sl(2)$. The $sl(2)$ doublet index is denoted $I,J,\ldots$ and the 4 of
$so(6)$ by a lower $A,B,\ldots$. A collective 8-dimensional Dirac
spinor index for 
$4\oplus\bar4$ is written $\alpha,\beta,\ldots$.
When spinor indices for $so(6)\oplus so(6)$ are introduced, we use
lower $A,B,\ldots$ for spinors in 4 under the first $so(6)$ and lower
$A',B',\ldots$ for the second.

\section{World-volume and background fields}

Utilising the $d=7$ Clifford algebra from \cite{Bao:2005}, reduced to
$d=6$, we may write the target space Bianchi identities as Fierz
identities and derive the following $\mathrm{dim}=0$ field strengths: 
\begin{equation}
\left\{ \begin{array}{lll}
F_{(2)}: && F_{\alpha I, \beta J} = 2 \varepsilon_{IJ} (\gamma^7)_{\alpha \beta}, \\
F_{(4)}: && F_{ab, \alpha I, \beta J} = -2 \varepsilon_{IJ} (\gamma_{ab})_{\alpha \beta}, \\
F_{(6)}: && F_{abcd, \alpha I, \beta J} = -2 \varepsilon_{IJ} (\gamma_{abcd} \gamma^7)_{\alpha \beta}, \\
H_{(3)}: && H_{a, \alpha I, \beta J} = 2 \varepsilon_{IJ} (\gamma_a \gamma^7)_{\alpha \beta}, \\
H_{(7)}: && H_{abcde, \alpha I, \beta J} = 2 \varepsilon_{IJ} (\gamma_{abcde})_{\alpha \beta}, \\
& & T_{\alpha I, \beta J}^a = 2 \varepsilon_{IJ} (\gamma^a)_{\alpha \beta}.
\end{array} \right.
\end{equation}
These field strengths obey the modified Bianchi identities
\begin{equation}
\begin{array}{lll}
&& dF_{(2)} = 0, \\
&& dF_{(4)} + F_{(2)} \wedge H_{(3)} = 0, \\
&& dF_{(6)} + F_{(4)} \wedge H_{(3)} = 0,
\end{array}
\end{equation}
\begin{equation}
\begin{array}{lll}
&& dH_{(3)} = 0, \\
&& dH_{(7)} - F_{(2)} \wedge F_{(6)} + \frac{1}{2} F_{(4)} \wedge F_{(4)} = 0.
\end{array}
\end{equation}
The 'maximal form' formalism relies on coupling background field strengths to the world volume of the object one wishes to model. In the current model we will do so for all background fields apart from the NS-NS 2-form potential $B_{(2)}$ since such a coupling would signal the possibility of ending strings on NS-branes; a non-existent scenario. The couplings are all of the basic form $f = da_{\mathrm{world volume}} - A_{\mathrm{background}}$ with additional terms introduced by the modified nature of the background Bianchi identities. These, together with the exclusion of $B_{(2)}$, completely fix the world-volume Bianchi identities 
\begin{equation}
\begin{array}{lll}
&& df_{(1)} = -F_{(2)}, \\
&& df_{(3)} = -F_{(4)} - f_{(1)} \wedge H_{(3)}, \\
&& df_{(5)} = -F_{(6)} - f_{(3)} \wedge H_{(3)},
\end{array}
\end{equation}
\begin{equation}
\begin{array}{lll}
&& dh_{(2)} = -H_{(3)}, \\
&& dh_{(6)} = -H_{(7)} - \frac{1}{2} f_{(1)} \wedge F_{(6)} + \frac{1}{2} f_{(3)} \wedge F_{(4)} -\frac{1}{2} f_{(5)} \wedge F_{(2)},
\end{array}
\end{equation}
where the Bianchi identities for the RR-fields are neatly summarised by
\begin{equation}
df + f \wedge H_{(3)} + F = 0 \label{bi_f}. 
\end{equation}
Stipulating further that these field strengths should be gauge invariant (again with the exclusion of the background 2-form potential from fields coupled to the world-volume) fixes the field strengths in terms of field potentials:
\begin{equation}
\begin{array}{rcl}
f &=& da - A + a \wedge H_{(3)}, \\
h_{(2)} &=& db_{(1)} - B_{(2)}, \\
h_{(6)} &=& db_{(5)} - \frac{1}{2} a_{(0)} \wedge F_{(6)} + \frac{1}{2} a_{(2)} \wedge F_{(4)} - \frac{1}{2} a_{(4)} \wedge F_{(2)},
\end{array} \label{wvfields}
\end{equation}
where we write
\begin{equation}
\begin{array}{rcl}
a &=& a_{(0)} + a_{(2)} + a_{(4)},\\
f &=& f_{(1)}+f_{(3)}+f_{(5)}
\end{array}
\end{equation}
as formal sums. 
%from which the exact expressions can be retrieved by fixing the form
%degree of the total expression. 
Further on we will write this sum as a structure in components of the differential forms and endow it with a symmetry larger than $so(6)$.

\section{Action and self-duality}

Maximal form models all share the same basic form of action:
\begin{equation}
S = \int d^6 \xi \sqrt{g} \lambda \left( 1 + \Phi(f) + (\star h_{(6)})^2 \right), \label{implaction}
\end{equation}
where $g$ is the determinant of the world-volume metric, $\lambda$ is a Lagrange multiplier and $\Phi$ is an, as of yet unknown, polynomial in all the world-volume field strengths apart from the maximal form $h_{(6)}$. This action gives us the implicit equations of motion 
\begin{eqnarray}
\lambda &:& 1 + \Phi(f) + (\star h_{(6)})^2 = 0 \label{eomlambda} \\
& & \Rightarrow \star h_{(6)} = -i \sqrt{1 + \Phi} = - iN, \nn \\
b_5 &:& d(\lambda \star h_{(6)}) = 0, \label{eomb5} \\
a &:& - d(\lambda \star q) + \lambda \star q \wedge H_{(3)} - \lambda (\star h_{(6)}) \pi F = 0, \label{eom_a}
\end{eqnarray}
where $q = \frac{\partial \Phi}{\partial f}$ and the minus sign has been chosen for the square root in (\ref{eomlambda}). The operator $\pi$ acts on some structure of differential forms, changing their sign based on form degree, and is defined by
\begin{eqnarray}
&&\pi(\omega_{(0)} + \omega_{(2)} + \omega_{(4)} + \omega_{(6)}) = \omega_{(0)} - \omega_{(2)} + \omega_{(4)} - \omega_{(6)}, \\
&&\pi(\omega_{(1)} + \omega_{(3)} + \omega_{(5)}) = - \omega_{(1)} + \omega_{(3)} - \omega_{(5)}.
\end{eqnarray}
This action carries a large number of excess degrees of freedom as compared to the A-model NS5-brane itself. These are reduced by demanding an implicit non-linear self-duality relation, between the fields, that equates the equation(s) of motion for $a$ with the Bianchi identity of $f$. A straightforward comparison between (\ref{eom_a}) and (\ref{bi_f}), together with the use of (\ref{eomlambda}) and (\ref{eomb5}), gives the following implicit self-duality:
\begin{equation}
\star q = \pi (\star h_{(6)}) f \Longrightarrow i \pi \star q = N f, \label{implsd}
\end{equation}
which in more explicit terms (for later us in the $\kappa$-variation) becomes
\begin{eqnarray}
&&q^i = -iN (\star f_{(5)})^i, \\
&&q^{ijk} = iN (\star f_{(3)})^{ijk},\\
&&q^{ijklmn} = -iN (\star f_{(1)})^{ijklmn}.
\end{eqnarray}

The generalised B-model of ref. \cite{Hitchin:2004ut} carries an
$so(6,6)$ symmetry, which in this model is broken by the twisted
duality operator $i \pi \star$ to $so(6) \oplus so(6)$. We would like
to realise this symmetry, using the introduced fields, on the
NS5-brane world-volume. Now, $so(6,6) \rightarrow so(6) \oplus so(6)$
implies $32 \rightarrow (4,4) \oplus (\bar{4},\bar{4})$ and $32'
\rightarrow (4,\bar{4}) \oplus (\bar{4},4)$, and further breaking to the
diagonal $so(6)$ turns $32$ and $32'$ into odd and even ($\pi$-twisted)
self-dual forms. Starting with the $so(6)$-fields given in
(\ref{wvfields}) we may use a parametrisation of the $\gamma$-matrices
in terms of $\sigma$-matrices (see Appendix A for further details) to
construct the fields 
\begin{equation}
\left\{ \begin{array}{rcl}
f_{AB} & = &  i f_i \sigma^i{}_{AB} + \frac{1}{3!} f_{ijk} \sigma^{ijk}{}_{AB} - i f_{ijklm} \sigma^{ijklm}{}_{AB} \\
&=& 2 \bar{k}_i \sigma^i{}_{AB} + \frac{1}{3} k_{ijk} \sigma^{ijk}{}_{AB}, \\
\tilde{f}^{AB} & = & -i f_i \bar{\sigma}^{iAB} + \frac{1}{3!} f_{ijk} \bar{\sigma}^{ijkAB} + i f_{ijklm} \bar{\sigma}^{ijklmAB} \\
&=& 2 k_i \bar{\sigma}^{iAB} + \frac{1}{3} \bar{k}_{ijk} \bar{\sigma}^{ijkAB},
\end{array} \right.
\end{equation}
where we have defined
\begin{equation}
\left\{ \begin{array}{rcl}
k_i & = & -\frac{i}{2} ( f_i + i(\star f)_i ) \\
\bar{k}_i & = & \frac{i}{2}(f_i - i(\star f)_i) \\
f_{ijk} & = & (k_{ijk} + \bar{k}_{ijk})
\end{array}, \right.
\end{equation}
$k_{ijk}$ and $\bar{k}_{ijk}$ being self-dual and anti-selfdual forms respectively (i.e. $\star k_{ijk} = i k_{ijk}$ etc.).
The fields $f$ and $\tilde f$ are twisted self-dual and anti-selfdual
under $i \pi \star$. The precise way $f_{(1)}$ and $f_{(5)}$ should
accompany $f_{(3)}$ to form these $so(6)\oplus so(6)$-covariant fields
is not obvious; these are the actual combinations that turn out to be
demanded by $\kappa$-symmetry.
%and furthermore, are the two components of $4 \oplus \bar{4}$
%respectively. 
They are lifted to $(4,4) \oplus (\bar{4},\bar{4})$
simply by discarding their parametrisation in terms of
$\sigma$-matrices and writing $f_{AB'}$ and $\tilde{f}^{A'B}$. 

Using this new covariance we may proceed to solve the implicit
self-duality (\ref{implsd}) through purely algebraic means. It is
clear that the self-duality relations demand a term $\tr(f \tilde{f})$
in $\Phi$, after which we assume that the remaining part of $\Phi$ is
$so(6,6)$-invariant (it is not {\it a priori} obvious that this has to
be true, but the actual calculation shows that it is). Furthermore
there is a single quartic invariant, and it has to be a sum of the
terms $\det f$, $\det \tilde{f}$, $\tr(f \tilde{f} f \tilde{f})$ and
$(\tr(f \tilde{f}))^2$. Whereas the generators in $so(6,6)$, outside
of $so(6)\oplus so(6)$, transform as $(6,6)$ and act on $f$,
$\tilde{f}$ as 
\begin{equation}
\begin{array}{lllll}
\delta_M f_{AB'} &=& M_{AC,B'D'} \tilde{f}^{D'C} &=& \frac{1}{2} \varepsilon_{B'D'E'F'} M_{AC}{}^{E'F'} \tilde{f}^{D'C}, \\
\delta_M \tilde{f}^{A'B} &=& M^{BC,A'D'} f_{CD'} &=& \frac{1}{2} \varepsilon^{BCEF}M_{EF}{}^{A'D'}f_{CD'},
\end{array}
\end{equation}
a straightforward calculation shows that the quartic invariant is
\begin{equation}
R(f,\tilde{f})=\det f+\det \tilde{f}+\frac{1}{2} \tr(f \tilde{f} f \tilde{f}) - \frac{1}{4} (\tr(f \tilde{f}))^2.
\end{equation}
The Ansatz, now a linear combination of $\tr(f \tilde{f})$ and $R(f,\tilde{f})$, is established to be $\Phi= -\frac{1}{8}\tr(f \tilde{f}) - \frac{1}{8^{2}}R(f,\tilde{f})$, i.e.
\begin{equation}
\Phi = -\frac{1}{8} \tr(f \tilde{f}) - \frac{1}{8^2} \left( \det f +
  \det \tilde{f} +\frac{1}{2} \tr (f \tilde{f}f \tilde{f}) -
  \frac{1}{4}\left(\tr(f \tilde{f})\right)^2\right). \label{eq:explPhi} 
\end{equation}
This implies the non-linear self-duality relations 
\begin{equation} \label{selfduality}
\begin{array}{rcl}
- Nf &=& - f - \frac{1}{8} \left((\det \tilde{f}) \tilde{f}^{-1} + f \tilde{f} f - \frac{1}{2} f \tr(f \tilde{f})\right), \\
N \tilde{f} &=& - \tilde{f} - \frac{1}{8} \left( (\det f)f^{-1} + \tilde{f} f\tilde{f} - \frac{1}{2} \tilde{f} \tr(f\tilde{f}) \right).
\end{array}
\end{equation}
Tracing the first of these equations with $\tilde{f}$, the second with $f$
and adding them together gives $R(f,\tilde{f})= -4 \tr(f
\tilde{f})$, so that $N=\sqrt{1+\Phi}=\sqrt{1-\frac{1}{16} \tr(f\tilde{f})}$. The equations then give 
$f\tilde{f}=\frac{1}{4} \tr(f\tilde{f}) =
4 (1-N^2)\mathbbm{1}$, and finally 
$\det f=-16(1-N)(1+N)^3$ and $\det
\tilde{f}=-16(1-N)^3(1+N)$. The two 
equations (\ref{selfduality}) contain the same information, i.e. are consistent. Especially noteworthy among these expressions is the product $f \tilde{f}$ which, by multiplying with one of the matrix inverses results in the explicit non-linear self-duality relation:
\begin{equation}
\tilde{f} = 4(1-N^{2}) f^{{-1}}.
\end{equation}
With this expression for the non-linear self-duality at hand, along with our definitions, we may produce a dictionary, Table 1, for the translation of objects of $so(6) \oplus so(6)$ and $so(6)$. 
\begin{table}[t]
\centering 
\begin{tabular}{|l|l|} \hline
$so(6) \oplus so(6) $ & $so(6)$ \\ \hline
$f_{AB}$ & $2 \bar{k}_i \sigma^i{}_{AB} + \frac{1}{3} k_{ijk} \sigma^{ijk}{}_{AB}$ \\ \hline
$\tilde{f}^{AB}$ & $2 k_i \bar{\sigma}^{iAB} + \frac{1}{3} \bar{k}_{ijk} \bar{\sigma}^{ijkAB}$ \\ \hline
$\det(f)$ & $(k^2)^2 - 8 r^{ij} k_i k_j + \frac{8}{3} r_{ij} r^{ij}$ \\ \hline
$\det(\tilde{f})$ & $(\bar{k}^2)^2 - 8 \bar{r}^{ij} \bar{k}_i \bar{k}_j + \frac{8}{3} \bar{r}_{ij} \bar{r}^{ij}$ \\ \hline
%$(\det f) (f^{-1})^{AB}$ & $\left( - k^2 k^i + 4 r^{ij} k_j \right) \bar{\sigma}^{iAB} + \frac{1}{3!}$ \\ \hline
%$(\det \tilde{f}) (\tilde{f}^{-1})_{AB}$ & $...$ \\ \hline
$\tilde{f} = 4(1-N^{2})f^{-1}$ & $ \left\{ \! \begin{array}{l}\bar{k}_i = \frac{1}{(1+N)^2} \left( k^2 k_i + r_{ij} k^j \right) \\  \bar{k}_{ijk} = -\frac{1}{(1+N)^2} \left( r_{[i}^{\; \; l} k_{jk]l} + 6 k_{[i} k^l k_{jk]l} - k^2 k_{ijk} \right) \end{array} \right.$ \\ \hline
\end{tabular}
\caption{Dictionary for translation between $so(6) \oplus so(6)$ and $so(6)$}
\end{table}

\section{$\kappa$-symmetry}

The explicit action given by (\ref{eq:explPhi}), and the associated
duality relation, must be compatible with $\kappa$-symmetry. In fact
$\kappa$-symmetry uniquely determines the form of self-duality on the
world-volume and can thus be used to solve the problem which was
solved using algebraic means in the previous section (such an approach
is found e.g. in ref. \cite{Cederwall:1998a}). Here we will progress
conventionally, beginning with the derivation of $\kappa$-variations
for our world-volume fields as dictated by the background fields and
gauge invariance. These variations are then inserted into the
constraint ($1.5$-order formalism) originating in the equation of
motion for the Lagrange multiplier ({\it i.e.}, everything within the
parenthesis) from the {\it implicit} action (\ref{implaction}). A
solution, $\kappa$, is then calculated which requires self-duality and
associated relations.

The major difference between this calculation and other proofs of
$\kappa$-symmetry for maximal form models is complexity. Given the
relatively large number of fields this is expected; less expected is
the lack of an apparent avenues for simplification. The
$\kappa$-variation does not exhibit the covariance of $f_{AB'},
\tilde{f}^{A'B}$ which arises only at the level of solution. Here, we
give only an outline of this lengthy calculation. 

We use superspace conventions, meaning derivatives act from the right,
in which the $\kappa$-variations of our world-volume fields become: 
\begin{equation}
\begin{array}{rcl}
\delta_\kappa f_i &=& -i_\kappa F_{(2)}, \\
\delta_\kappa f_{ijk} &=& -i_\kappa F_{(4)} - f_{(1)} \wedge i_\kappa H_{(3)}, \\
\delta_\kappa f_{ijklm} &=& -i_\kappa F_{(6)} - f_{(3)} \wedge i_\kappa H_{(3)}, \\
\delta_\kappa h_{ijklmn} &=& -i_\kappa H_{(7)} - \frac{1}{2} f_{(1)}
\wedge i_\kappa F_{(6)} + \frac{1}{2} f_{(3)} \wedge i_\kappa F_{(4)}
- \frac{1}{2} f_{(5)} \wedge i_\kappa F_{(2)}, \\ 
\delta_\kappa g_{ij} &=& 4 E_{(i,| \alpha I |} (\gamma_{j)})^{\alpha
  I}_{\; \beta J} \kappa^{\beta J}. 
\end{array}
\end{equation}
The constraint  whose invariance we wish to prove,
\begin{equation}
\Psi = 1 + \Phi(f) + (\star h_{(6)})^{2},
\end{equation}
transforms as:
\begin{equation}
\begin{array}{rcl}
\delta_\kappa \Psi &=& q^i \delta_\kappa f_i - \frac{1}{2} f^i q^j
\delta_\kappa g_{ij} \\ 
 &+& \frac{1}{3!} q^{ijk} \delta_\kappa f_{ijk} - \frac{1}{4} f^i_{\;
   kl} q^{jkl} \delta_\kappa g_{ij} \\ 
&+& \frac{1}{5!} q^{ijklm} \delta_\kappa f_{ijklm} - \frac{1}{2 \cdot
  4!} q^i_{\; klmn} k^{jklmn} \delta_\kappa g_{ij} \\ 
&+& \frac{2}{6!} h^{ijklmn} \delta_\kappa h_{ijklmn} - \frac{1}{5!}
h^i_{\; klmnp} h^{jklmnp} \delta_\kappa g_{ij}, \label{deltapsi}
\end{array}
\end{equation}
and produces a result on the form
\begin{equation}
\delta_{\kappa} \Psi = E_{i,\alpha} (M^{i})^{\alpha}{}_{\beta} \kappa^{\beta}. 
\end{equation}
The matrix $M$ is a complex expression of field strengths and
$\sigma$-matrices (or, as this calculation can be performed in the
representation $8$ of $\mathrm{so}(6)$, $\gamma$-matrices) whereas
$\kappa$ exhibits the covariance of $f, \tilde{f}$:  
\begin{equation}
\kappa = \frac{1}{2} \left[ \underbrace{\left(\begin{array}{cc}
        \mathbbm{1}^A{}_{B} & 0 \\ 0 &
        \mathbbm{1}_A{}^{B} \end{array}\right)}_{\mathbbm{1}} +
  \underbrace{\frac{1}{N} \left(\begin{array}{cc}\mathbbm{1}^A{}_{B} &
        \frac{i}{2} \tilde{f}^{AB} \\ \frac{i}{2} f_{AB} &
        -\mathbbm{1}_A{}^{B} \end{array}\right)}_{\Gamma} \right]
\zeta. \label{projkappa}
\end{equation}
As usual, $\Gamma^{2} = \mathbbm{1}$. This is seen by using the
self-duality from the previous 
section in the form $f\tilde{f}=
\frac{1}{2}(N^2-1)\mathbbm{1}$, which implies 
%\begin{equation}
%\left( \frac{1}{2} \left[ \mathbbm{1} + \Gamma \right] \right)^{2} =
%\frac{1}{2} \left[ \mathbbm{1} + \Gamma \right] . 
%\end{equation}
that $\frac{1}{2}(\mathbbm{1}\pm\Gamma)$ are projection matrices.
The actual form of $\Gamma$ was actually guessed using input from the
case where the field strength only contains a 3-form
\cite{Cederwall:1998a,Bao:2006}, together with the observation that it
should respect $so(6)\oplus so(6)$.
This solution renders $\delta_{\kappa} \Psi = 0$, modulo terms that
vanish when the self-duality constraints are imposed, and thus the model is
$\kappa$-symmetric.

Although the projection on $\kappa$ is manifestly $so(6)\oplus
so(6)$-covariant, neither the action nor the $\kappa$-variations of
the fields are. Thus, the only local symmetry we can rely on in the
calculation is $so(6)$. This makes the number of terms to check quite
large,  and the calculation becomes long and cumbersome. We have not
been able to find a more efficient way than to decompose $M$ and
$\Gamma$ in terms of 6-dimensional $\gamma$-matrices, using the
explicit
parametrisation of $f$ and $\tilde f$, as well as the less covariant
fields occurring in the $\kappa$-variation (\ref{deltapsi}), 
in terms of the vector $k_i$ and
the linearly self-dual $k_{ijk}$ given in Table 1 together with the
projected $\kappa$ parameter of eq. (\ref{projkappa}), and check all terms.

\section{Pure spinors and self-dual odd forms}

When we have actions with the complete set of fields coupling to
background tensors as above, the world-volume degrees of freedom are
reduced by some self-duality condition. This will also be the case on
the NS5-brane of the A-model, where it will be natural to describe the
generalised complex structure in terms of self-duality.

Six dimensions is special, this is the only case where purity of a
spinor can be expressed as (non-linear) self-duality of an even/odd
form. The counting is trivial, both a pure spinor and a self-dual
even/odd form has 16 independent components. The rest of the
correspondence can be seen as follows. 

Let us first examine the case
where $\varphi$ is a 3-form. The purity condition says that 
$\iota_v \varphi = 0 = \varphi \wedge \eta$ (the $Spin(6,6)$
$\gamma$-matrices are 
$\Gamma_i \cdot \varphi = \iota_i \varphi$, $\Gamma^i \cdot \varphi =
\varphi \wedge e^i$),  
for a 3-dimensional subspace of
vectors $v$ and a three-dimensional subspace of 1-forms $\eta$
fulfilling $\iota_v \eta = 0$, {\it i.e.}, $ \Gamma(V) \cdot \varphi = 0$ for an
isotropic subspace of $V$'s.   
Given a linearly self-dual 3-form $h$ with $i{\star}h=h$, and
$r_{ij}=\frac{1}{2} h_i{}^{kl}h_{jkl}$, we can construct projection matrices
$P_\pm=\frac{1}{2}(\mathbbm{1} \pm \frac{r}{\rho})$, where $\pm \rho$
are the roots to 
$\rho^2=\frac{1}{6} \tr r^2$, as long as the quartic invariant
$\tr r^2$ is non-zero. We can then write $\varphi=P_+h$ (as usual, it
suffices to act on one index, since
$r_{[i}{}^lh_{jk]l}=r_{i}{}^lh_{jkl}$). This 3-form $\varphi$ is
annihilated by $\iota_v \cdot$ and by $\cdot \wedge \eta$ for $v=P_-
w$, $\eta=P_+ \nu$. 

%[This is not the complete story. Some further condition has to be put
%on $h$, so that the pure spinor becomes non-singular in the sense
%$\varphi \cdot \bar{\varphi} = \varphi \wedge \sigma \bar{\varphi} =
%0$ (this is the 
%symplectic spinor scalar product). Otherwise the construction above
%includes e.g. real pure spinors for which this product vanishes: take
%$h_{123}=1$, $h_{456}=-i$, then $P_+$ projects on the first three
%components. $\varphi_{123}=1$, the rest zero, is certainly a pure
%spinor, although a ``singular'' one and not a candidate $\Omega$.]
 
We can think of $P_\pm$ as projections on holomorphic and
anti-holomorphic components, adapted to a complex structure (not yet
generalised) defined by $h$. This makes the rest, the inclusion of a
1-form and a 5-form, somewhat less technical. Denote the 3-form above
by $\Omega$, and think about it as a holomorphic 3-form.
The deformation from $\varphi=\Omega$ to a $\varphi$ containing a
1-form and a 5-form can be achieved by a transformation in
$Spin(6,6)\times\mathbb{C}$. 
%Essentially, the transformation
%parameters are
%$i_\xi\Omega$ and ${\star}($, where $w$ is a holomorphic vector
%and $\bar\nu$ an anti-holomorphic 
It is however easier to find $\varphi$ by trial and error.
It turns out that 
\begin{equation}
\varphi=\xi + \Omega + \iota_{\bar{\zeta}} \bar{\Omega} \wedge \xi + \star \bar{\zeta},
\end{equation}
where $\xi$ is a holomorphic 1-form and $\bar{\zeta}$ an
anti-holomorphic vector density. This pure spinor only has form
degrees (1,0), (3,0), (1,2) and (3,2) with respect to the complex
structure defined by $\Omega$. 
It is annihilated by an isotropic subspace,
$\iota_v \varphi + \varphi \wedge \eta = 0$, with
\begin{eqnarray}
v &=&\bar{w} - \star( \bar{\Omega} \wedge \xi \wedge \nu),\\
\eta &=& \nu - \iota_{\bar w} \iota_{\bar{\zeta}} \bar{\Omega},
\end{eqnarray}
where $\bar{w}$ is an anti-holomorphic vector and $\nu$ a holomorphic
1-form density. We have defined duality (without raising or lowering
indices) with $\varepsilon$, and $\Omega_{abc}=\varepsilon_{abc}$.

The solution can be parametrised in terms of a linearly (twisted)
self-dual form $h$, which up to rescaling is the component $f$ of the
previous sections. One
convenient choice is
\begin{eqnarray}
f &=& 4(1+N)h, \\
\tilde{f} &=& -16 (1+N)R(h,0)h^{-1},
\end{eqnarray}
with $R(h,0)=\det h=-\frac{1}{16}\frac{1-N}{1+N}$.
Given a linearly self-dual $h$, one can construct a pure spinor
$(\varphi,\tilde{\varphi})$.
Taking
\begin{eqnarray}
\varphi &=&c(N) h, \\
\tilde{\varphi} &=& \pm c(N) \sqrt{R(h,0)}h^{-1}
\end{eqnarray} 
for some function $c$
defines a pure odd form $(\varphi,\tilde{\varphi})$ with
$R(\varphi,\tilde{\varphi}) = 0$. Note that the anti-selfdual part of the
pure spinor scales as $h$, while the anti-selfdual part of the
field strength above starts with a term scaling as $h^3$. The
relations above can of course be used to obtain a direct relation
between the field strength and the pure spinor. 
%but why it looks
%exactly as it does is not obvious...

\appendix

\section{$\sigma$-matrix algebra}

In lifting $so(6)$-covariant expressions to $so(6) \oplus so(6)$ we have used a parametrisation of the Clifford algebra
\begin{equation}
\{ \gamma^a, \gamma^b \} = -2 \delta^{ab},
\end{equation}
in terms of $\sigma$-matrices,
\begin{equation}
(\gamma^a)^\alpha_{\; \beta} = \left[ \begin{array}{cc} 0 & \bar{\sigma}^{aAB} \\ \sigma^a{}_{AB} & 0 \end{array}\right],
\end{equation}
that reduces this algebra to:
\begin{equation}
\sigma^{(a} \bar{\sigma}^{b)} = -2 \delta^{ab}.
\end{equation}
We also define $\gamma^7$ which transforms symmetric matrices to anti-symmetric ones in our calculations,
\begin{equation}
(\gamma^7)^\alpha{}_{\beta} = \frac{-1}{6!} \varepsilon_{abcdef} (\gamma^{abcdef})^\alpha{}_{\beta}= -(\gamma^1 \gamma^2 \gamma^3 \gamma^ 4 \gamma^5 \gamma^ 6)^\alpha{}_\beta =\left[ \begin{array}{cc}i \mathbbm{1}^A_{\; B} & 0 \\0 & -i \mathbbm{1}_A^{\; B}\end{array}\right].
\end{equation}
These matrices are explicitly defined by
\begin{equation}
\left\{ \begin{array}{cccc}
\sigma^{1,2,3} & = & \frac{1}{\sqrt{2}} (A_x + B_x) & , x = 1,2,3 \\
\sigma^{4,5,6} & = & \frac{1}{\sqrt{2}} (A_x - B_x) & , x = 1,2,3
\end{array}, \right.
\end{equation}
with the following basis:
\begin{eqnarray}
A_1 = \left[\begin{array}{cccc}0 & 0 & 1 & 0 \\0 & 0 & 0 & -1 \\-1 & 0 & 0 & 0 \\0 & 1 & 0 & 0\end{array}\right] &,& B_1 = \left[\begin{array}{cccc}0 & i & 0 & 0 \\-i & 0 & 0 & 0 \\0 & 0 & 0 & -i \\0 & 0 & i & 0\end{array}\right], \nn \\
A_2 = \left[\begin{array}{cccc}0 & 1 & 0 & 0 \\-1 & 0 & 0 & 0 \\0 & 0 & 0 & 1 \\0 & 0 & -1 & 0\end{array}\right] &,& B_2 = \left[\begin{array}{cccc}0 & 0 & i & 0 \\0 & 0 & 0 & i \\-i & 0 & 0 & 0 \\0 & -i & 0 & 0\end{array}\right], \\
A_3 = \left[\begin{array}{cccc}0 & 0 & 0 & 1 \\0 & 0 & 1 & 0 \\0 & -1 & 0 & 0 \\-1 & 0 & 0 & 0\end{array}\right] &,& B_3 = \left[\begin{array}{cccc}0 & 0 & 0 & i \\0 & 0 & -i & 0 \\0 & i & 0 & 0 \\-i & 0 & 0 & 0\end{array}\right]. \nn 
\end{eqnarray}
The dual, or complex conjugate, matrices are given calculated by:
\begin{equation}
\bar{\sigma}^{aAB} = (\sigma^*)^{aAB} = \frac{1}{2}\varepsilon^{ABCD}\sigma^a_{\; CD}.
\end{equation}
The $\sigma$-matrices dualise as:
\begin{equation}
\begin{array}{l}
\left\{ \begin{array}{ccc}
\star \sigma^{(1)} & = & i\sigma^{(5)} \\
\star \sigma^{(3)} & = & -i \sigma^{(3)} \\
\star \sigma^{(5)} & = & i \sigma^{(1)}
\end{array} \right.
 \Rightarrow 
\left\{ \begin{array}{ccc}
\star \bar{\sigma}^{(1)} & = & -i\bar{\sigma}^{(5)} \\
\star \bar{\sigma}^{(3)} & = & i\bar{\sigma}^{(3)} \\
\star \bar{\sigma}^{(5)} & = & -i\bar{\sigma}^{(1)}
\end{array} \right. \\
\left\{ \begin{array}{ccc}
\star \sigma^{(2)} & = & -i\sigma^{(4)} \\
\star \sigma^{(4)} & = & i\sigma^{(2)} \\
\star \mathbbm{1}^A{}_{B} & = & i \sigma^{(6)A}{}_{B}
\end{array} \right.
 \Rightarrow 
\left\{ \begin{array}{ccc}
\star \bar{\sigma}^{(2)} & = & i\bar{\sigma}^{(4)} \\
\star \bar{\sigma}^{(4)} & = & -i\bar{\sigma}^{(2)} \\
\star \mathbbm{1}_A{}^{B} & = & -i \bar{\sigma}^{(6)}{}_{A}{}^{B}
\end{array} \right.
\end{array}.
\end{equation}

\bibliographystyle{JHEP}
\bibliography{ns5brane}

\end{document}